# PQCMC: Post-Quantum Cryptography McEliece-Chen Implicit Certificate Scheme

Abel C. H. Chen
*Information & Communications Security Laboratory,
Chunghwa Telecom Laboratories*
Taoyuan, Taiwan
ORCID: 0000-0003-3628-3033; Email: chchen.scholar@gmail.com

*Abstract*—In recent years, the elliptic curve Qu-Vanstone (ECQV) implicit certificate scheme has found application in security credential management systems (SCMS) and secure vehicle-to-everything (V2X) communication to issue pseudonymous certificates. However, the vulnerability of elliptic-curve cryptography (ECC) to polynomial-time attacks posed by quantum computing raises concerns. In order to enhance resistance against quantum computing threats, various post-quantum cryptography methods have been adopted as standard (e.g. Dilithium) or candidate standard methods (e.g. McEliece cryptography), but state of the art has proven to be challenging to implement implicit certificates using lattice-based cryptography methods. Therefore, this study proposes a post-quantum cryptography McEliece-Chen (PQCMC) based on an efficient random invertible matrix generation method to issue pseudonymous certificates with less computation time. The study provides mathematical models to validate the key expansion process for implicit certificates. Furthermore, comprehensive security evaluations and discussions are conducted to demonstrate that distinct implicit certificates can be linked to the same end entity. In experiments, a comparison is conducted between the certificate length and computation time to evaluate the performance of the proposed PQCMC. This study demonstrates the viability of the implicit certificate scheme based on PQC as a means of countering quantum computing threats.

*Keywords—implicit certificate, post-quantum cryptography, McEliece cryptography, code-based cryptography, ECQV*

## I. INTRODUCTION

Due to the development of quantum computing and Shor's algorithm [1], numerous mainstream asymmetric cryptography techniques (such as RSA and elliptic-curve cryptography (ECC)) are susceptible to polynomial-time attacks. Consequently, the US National Institute of Standards and Technology (NIST) has initiated a call for proposals in the domain of post-quantum cryptography (PQC) methods. Various categories of PQC methods have emerged, including lattice-based cryptography, code-based cryptography, hash-based cryptography, multivariate-based cryptography, and supersingular elliptic curve isogeny cryptography [2]. Furthermore, lattice-based cryptography has emerged as one of the leading PQC approaches, offering efficient operations for encryption, decryption, signature, and verification [2].

For generating implicit certificates, the elliptic curve Qu-Vanstone (ECQV) technique is a widely used approach to achieve pseudonymous certificates within security credential management systems (SCMS) and secure vehicle-to-everything (V2X) communication [3],[4]. In efforts to enhance the security of SCMS and V2X communication, Shim conducted a survey to consider potential standard post-quantum cryptography (PQC) methods (such as Dilithium and Rainbow) and compared the performance of PQC methods and ECC methods [5]. In the realm of issuing implicit certificates for SCMS and V2X communication, Bindel and McCarthy tried to design emulating ECQV methods utilizing Falcon and Dilithium, and they examined common failed patterns based on lattice-based cryptography methods [6]. It has proven to be challenging to implement implicit certificates using lattice-based cryptography methods.

Therefore, this study designs an ECQV-like code-based PQC implicit certificate scheme for the issuance of implicit certificates in SCMS and V2X communication. A post-quantum cryptography McEliece-Chen (PQCMC) is proposed, which offers implicit certificates based on the principles of McEliece cryptography. The contributions of this study are summarized and emphasized as follows.

- The proposed PQCMC implicit certificate scheme enables the provision of pseudonymous certificates without necessitating signatures.

- Mathematical models are given to prove the process of key expansion for implicit certificates.

- An efficient method for generating a random invertible matrix, with a time complexity of $O(n)$, is proposed to support the proposed PQCMC scheme with less computation time.

The structure of this manuscript include five sections. Section II presents the background of ECC, ECQV, and McEliece cryptography. Section III illustrates the proposed PQCMC and provides the proof and applications. Section IV obtains the evaluation and comparison of the proposed PQCMC. Finally, Section V concludes the contributions of this study and discusses the future work.

## II. BACKGROUND

In this section, the theories of ECC and elliptic curve digital signature algorithm (ECDSA) are presented in Subsection II.A and Subsection II.B. For illustration of implicit certificates, ECQV implicit certificate scheme and the implicit certificate schemes in security credential management system (SCMS) are described in Subsection II.C and Subsection II.D, respectively. Lastly, McEliece cryptography and McEliece-based digital signature scheme are shown in Subsection II.E and Subsection II.F.

### A. Elliptic Curve Cryptography

An elliptic curve (EC) is defined as **Eq. (1)**, and the coordinate of an EC point is denoted as $(x, y)$. The values of a coefficient $\alpha$, a constant $\beta$, and a prime modulus $n$ in **Eq. (1)** are defined in the specification [7] for various security levels. For key generation, a private key $a$ is randomly generated, and the public key $A$ $(A_x, A_y)$ can be calculated based on a base point $G$ $(G_x, G_y)$ by $A = aG$; the EC point addition and EC point doubling can been found in [7],[8].

$$y^2 = x^3 + \alpha x^2 + \beta \pmod{n}. \tag{1}$$

For compression, an EC point can be compressed to only include a tag and the *x*-coordinate, and the tag can represent whether the last bit of *y*-coordinate is one or zero. For decompression, the value of the *y*-coordinate can be calculated by **Eq. (1)** in accordance with the tag and the *x*-coordinate. For instance, the EC of NIST P-256 is selected as one of standard ECs for SCMS [3]. An uncompressed EC point has a length of 65 bytes, which includes a tag (one byte), a *x*-coordinate (32 bytes), and a *y*-coordinate (32 bytes). Moreover, a compressed EC point has a length of 33 bytes, including only a tag (one byte) and a *x*-coordinate (32 bytes). Consequently, utilizing a compressed EC point can save thirty-two bytes.

*B. Elliptic Curve Digital Signature Algorithm*

For the illustration of ECDSA, an ECC-based private key *a* and the public key is *A* are selected in this subsection. The hash of the to-be-signed message *m* is represented as *h* by a hash function *H*(*m*) (e.g. Secure Hash Algorithm-256 (SHA-256)). In the process of signature generation, an integer number *r* is randomly generated, and the EC point *R* ($x_R$, $y_R$) is determined by *R* = *rG*. The value of an integer number *s* can be calculated by **Eq. (2)**, and the signature (*R*, *s*) can be obtained [9],[10].

$$s = [(h + a x_R) / r] \pmod{n}. \tag{2}$$

For signature verification, the EC point *Z* can be calculated using **Eqs. (3), (4), (5),** and **(6)**. Furthermore, if the EC point *Z* is equals to the EC point *R*, the signature verification passes. The mathematical proof for this assertion is represented by **Eq. (7)** [9],[10].

$$w = 1 / s = [r / (h + a x_R)] \pmod{n}. \tag{3}$$

$$u = hw = h / s = hr / (h + a x_R). \tag{4}$$

$$v = x_R w = x_R / s = x_R r / (h + a x_R). \tag{5}$$

$$Z = uG + vA = uG + vaG = (u + va)G. \tag{6}$$

$$u + va = [hr / (h + a x_R)] + [a x_R r / (h + a x_R)] = r. \tag{7}$$

For the example of NIST P-256, The length of the signature (*R*, *s*) is 97 bytes with an uncompressed EC point *R* (65 bytes) and the integer number *s* (32 bytes). Furthermore, the length of the signature can be reduced to 65 bytes [4] by utilizing a compressed EC point *R* (33 bytes) mentioned in Subsection II.A.

*C. Elliptic Curve Qu-Vanstone*

In the description of ECQV, End Entity 1 (EE 1) possesses its ECC-based private key *a*, public key *A*, and information *E* in this subsection. Furthermore, a certificate authority (CA) has its ECC-based private key *c* and public key *C* which are used to generate the implicit certificate of EE 1 ($C_E$) based on the encoding of the reconstruction point (*P*) and the information of EE 1 (*E*). **Fig. 1** illustrates the ECQV process for generating the implicit certificate $C_E$ by the CA. The process and proof of ECQV are presented in the following subsections.

*1) The process of ECQV:* Firstly, EE 1 sends its public key *A* and information *E* to the CA. Subsequently, the CA generates a random number *r* which is then combined with the public key *A* to derive the reconstruction point *P*. The implicit certificate of EE 1 ($C_E$) can be encoded using the reconstruction point (*P*) and EE 1's information (*E*). Furthermore, the hash value (*h*) of the implicit certificate of EE 1 ($C_E$) is obtained through the application of the hash function *H*($C_E$). The private key reconstruction data *b* can be determined using the hash value *h*, the random value *r*, and the CA's private key *c*. Then the CA sends the value of *b* along with the implicit certificate $C_E$ to EE 1. Finally, EE 1 calculates the hash value (*h*) of the implicit certificate ($C_E$) which can be combined with the private key *a* and the value of *b* to generate the private key *q*. For retrieving the public key of EE 1 (*Q*), the hash value (*h*), the reconstruction point (*P*), and the public key of the CA (*C*) are utillized, so the issuer of the implicit certificate can be verified [11]-[13].

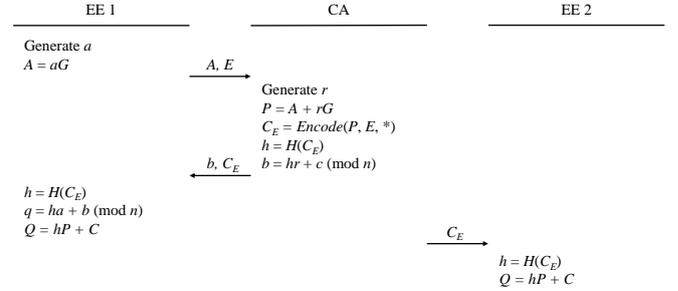

Fig. 1. The process of ECQV

*2) The proof of ECQV:* The detailed proof is illustrated in **Eq. (8)**. The EC point *qG* is equivalent to the EC point *Q*. Thus, EE 1 can utilize the private key *q* to sign a message with a corresponding signature, and EE 2 can use the public key *Q* relying on the implicit certificate $C_E$ and the CA's public key *C* to verify the signature [11]-[13].

$$\begin{aligned} qG &= (ha + b)G \\ &= (ha + hr + c)G \\ &= h(aG + rG) + cG \\ &= h(A + rG) + C \\ &= hP + C = Q. \end{aligned} \tag{8}$$

*D. ECQV Implicit Certificate Scheme in SCMS*

The structure of certificate in SCMS, as defined in the IEEE 1609.2 standard [3], includes version, type, issuer, to-be-signed certificate, and signature, as depicted in **Fig. 2**. The type indicates whether the certificate is explicit or implicit. An explicit certificate has a verification key (an EC point) within the verify key indicator (VKI) and the signature signed by a CA. Furthermore, an implicit certificate has a reconstruction value (an EC point) within the VKI, but it does not contain a signature [3].

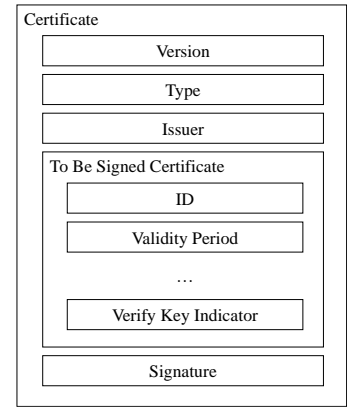

Fig. 2. The structure of certificate in SCMS

For the case of the certificate utilizing NIST P-256, the total length of a VKI and a signature with compressed EC points in an explicit certificate is 98 bytes (33 bytes for the VKI and 65 bytes for the signature), and the total length of a VKI and a signature with compressed EC points in an implicit certificate is 33 bytes (33 bytes for the VKI and 0 bytes for the signature). Therefore, the implicit certificate saves 65 bytes.

*E. McEliece Cryptography*

This subsection introduces McEliece cryptography for encrypting plaintext $m$ and decrypting ciphertext $z$. During key generation, a private key (i.e. $\{K_1, K_2, K_3\}$) is constructed comprising a scrambler $K_1$ (with dimensions $\zeta_1 \times \zeta_1$), a generator matrix $K_2$ (specifically a $\zeta_1 \times \zeta_2$ encoder matrix), and a permutation matrix $K_3$ (with dimensions $\zeta_2 \times \zeta_2$). Furthermore, the public key $L$ (with dimensions $\zeta_1 \times \zeta_2$) is obtained as the product of these three matrices (i.e. $L = K_1 K_2 K_3$). For decoding and error-correcting, a decoder matrix $K_4$ (with dimensions $\zeta_2 \times \zeta_1$) and an error-detector matrix $K_5$ are generated based on the generator matrix $K_2$ [14],[15].

To perform encryption, an encryption function $e(m, L)$ defined as **Eq. (9)** is used to encrypt plaintext $m$ using the public key $L$ and a random number $r$ resulting in ciphertext $z$ [14],[15].

$$e(m, L) = mL + r = mK_1 K_2 K_3 + r = z. \tag{9}$$

To perform decryption, a decryption function $d(z, \{K_1, K_2, K_3\})$ defined as **Eq. (10)** is used to decrypt plaintext $z$ using the private key $\{K_1, K_2, K_3\}$ resulting in plaintext $m$ (with dimensions $1 \times \zeta_1$). Furthermore, an error-correcting function $f(\alpha, K_5)$ is utilized based on the error-detector matrix $K_5$ to detect and remove the random number $r$. The decoder matrix $K_4$ can be applied to decode the encoded message produced by the encoder matrix $K_2$ (i.e. $xK_2 K_4 = x$) [14],[15].

$$
\begin{aligned}
d(z, \{K_1, K_2, K_3\}) &= f(zK_3^{-1})K_4 K_1^{-1} \\
&= f((mL + r)K_3^{-1})K_4 K_1^{-1} \\
&= f((mK_1 K_2 K_3 + r)K_3^{-1})K_4 K_1^{-1} \\
&= f(mK_1 K_2 K_3 K_3^{-1} + rK_3^{-1})K_4 K_1^{-1} \\
&= mK_1 K_2 K_4 K_1^{-1} \\
&= mK_1 K_1^{-1} = m.
\end{aligned}
\tag{10}
$$

*F. McEliece-based Digital Signature Scheme*

This subsection presents the McEliece-based digital signature scheme, which involves a private key $\{K_1, K_2, K_3\}$, a public key $L$, a decoder matrix $K_4$ based on the generator matrix $K_2$.

To generate the signature $s$ (with dimensions $\zeta_2 \times 1$) of the message $m$ (with dimensions $\zeta_1 \times 1$), the signature function $S(m, \{K_1, K_2, K_3\})$ is performed using the private key $\{K_1, K_2, K_3\}$ as defined in **Eq. (11)**. Furthermore, the verification function is executed with the utilization of the public key $L$ as defined in **Eq. (12)**, to verify the signature $s$ [16],[17].

$$S(m, \{K_1, K_2, K_3\}) = K_3^{-1} K_4 K_1^{-1} m = s. \tag{11}$$

$$
\begin{aligned}
V(s, L) &= Ls \\
&= K_1 K_2 K_3 K_3^{-1} K_4 K_1^{-1} m \\
&= K_1 K_2 K_4 K_1^{-1} m \\
&= K_1 K_1^{-1} m = m.
\end{aligned}
\tag{12}
$$

### III. THE PROPOSED METHODS

This section proposes a PQCMC implicit certificate scheme in Subsection III.A. The proof and security evaluation of the proposed PQCMC are discussed in Subsection III.B and Subsection III.C. Furthermore, an efficient method for generating a random invertible matrix is proposed in Subsection III.D.

*A. The Process of the Proposed PQCMC*

In the illustration of PQCMC, EE 1 possesses its McEliece-based private key $\{K_{1,(E)}, K_{2,(E)}, K_{3,(E)}\}$, public key $L_{(E)}$, and information $E$ as detailed in this subsection. Furthermore, a certificate authority (CA) holds its McEliece-based private key $\{K_{1,(CA)}, K_{2,(CA)}, K_{3,(CA)}\}$ and public key $L_{(CA)}$ which are utilized to generate the implicit certificate of EE 1 ($C_E$) based on the encoding of the McEliece-based reconstruction value ($B$) and the information of EE 1 ($E$). **Fig. 3** illustrates the PQCMC process for generating the implicit certificate $C_E$ by the CA. The PQCMC process is presented as follows.

Firstly, EE 1 generates its private key $\{K_{1,(E)}, K_{2,(E)}, K_{3,(E)}\}$. The corresponding public key $L_{(E)}$ (i.e. $K_{1,(E)} K_{2,(E)} K_{3,(E)}$) can be determined and then sent to the CA. Subsequently, the CA generates a random number $r$ and a random invertible matrix $M_r$ (with dimensions $\zeta_1 \times \zeta_1$) based on $r$. The random invertible matrix $M_r$ can be signed by the CA using the signature function $S(M_r, \{K_{1,(CA)}, K_{2,(CA)}, K_{3,(CA)}\})$ (i.e. $K_{3,(CA)}^{-1} K_{4,(CA)} K_{1,(CA)}^{-1} m$) to produce $T$. Furthermore, the matrix $T$ can be encrypted by EE 1's public key and the encryption function $e(T, L_{(E)})$ (i.e. $TL_{(E)}$) to create $B$ which serves as a reconstruction value. The implicit certificate of EE 1 ($C_E$) includes the reconstruction value $B$, EE 1's information $E$, and other necessary elements. Importantly, no signature is required for the implicit certificate. Finally, the CA transmits the random number $r$ along with EE 1's certificate $C_E$ to EE 1.

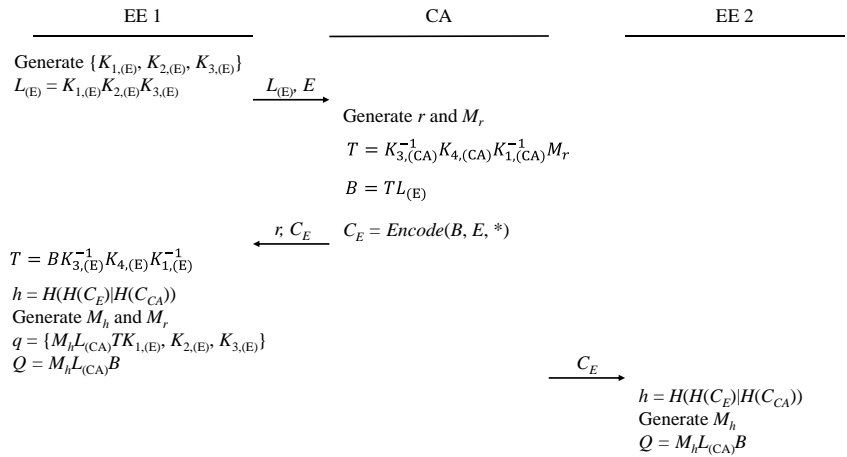

Fig. 3. The process of PQCMC

To generate the updated private key, EE 1 can decrypt the reconstruction value $B$ using its private key (i.e. $\{K_{1,(E)}, K_{2,(E)}, K_{3,(E)}\}$) with the decryption function $d(B, \{K_{1,(E)}, K_{2,(E)}, K_{3,(E)}\})$ to obtain the matrix $T$. The hash value of the concatenation of $H(C_E)$ and $H(C_{CA})$ is calculated as $h$ based on EE 1's certificate $C_E$ and the CA's certificate $C_{CA}$ for generating a random invertible matrix $M_h$ (with dimensions $\zeta_1 \times \zeta_1$). Finally, the updated private key of EE 1 can be represented as $q$ (i.e. $\{M_h L_{(CA)} T K_{1,(E)}, K_{2,(E)}, K_{3,(E)}\}$).

To generate the updated public key, each device can receive EE 1's certificate $C_E$ and extract the reconstruction value $B$ from $C_E$. The hash value $h$ and the random invertible matrix $M_h$ can be calculated following the procedure described in the previous paragraph. Consequently, the updated public key of EE 1 can be expressed as $Q$ (i.e. $M_h L_{(CA)} B$) utilizing the hash value $h$ and the CA's public key $L_{(CA)}$.

### B. Proof

This subsection proves the proposed PQCMC based on mathematical models. The plaintext $m$ can be signed by the CA's private key, resulting in the signature $K_{3,(CA)}^{-1} K_{4,(CA)} K_{1,(CA)}^{-1} m$ as described in **Eq. (11)**. This signature can then be verified using the CA's public key, allowing the retrieval of $m$ (as shown in **Eq. (13)**). Furthermore, the plaintext $m$ can be encrypted by EE 1's public key, leading to $mL_{(E)}$, as specified in **Eq. (9)**. Subsequently, decryption using EE 1's private key, as outlined in **Eq. (10)**, restores the original plaintext $m$ (as demonstrated in **Eq. (14)**).

$$L_{(CA)} K_{3,(CA)}^{-1} K_{4,(CA)} K_{1,(CA)}^{-1} m \\ = K_{1,(CA)} K_{2,(CA)} K_{3,(CA)} K_{3,(CA)}^{-1} K_{4,(CA)} K_{1,(CA)}^{-1} m = m. \quad (13)$$

$$mL_{(E)} K_{3,(E)}^{-1} K_{4,(E)} K_{1,(E)}^{-1} \\ = m K_{1,(E)} K_{2,(E)} K_{3,(E)} K_{3,(E)}^{-1} K_{4,(E)} K_{1,(E)}^{-1} = m. \quad (14)$$

The updated EE 1's public key $Q$ (i.e. $M_h L_{(CA)} B$) can be derived from the updated EE 1's private key $q$ (i.e. $\{M_h L_{(CA)} T K_{1,(E)}, K_{2,(E)}, K_{3,(E)}\}$). **Eq. (15)** proves that the multiplication of the private key $\{M_h L_{(CA)} T K_{1,(E)}, K_{2,(E)}, K_{3,(E)}\}$ (i.e. $M_h L_{(CA)} T K_{1,(E)} K_{2,(E)} K_{3,(E)}$) is equivalent to the public key $M_h L_{(CA)} B$. Furthermore, the inverse matrix of the first matrix in the private key can be computed using **Eq. (16)**, enabling encryption and signing operations based on the private key.

$$\begin{aligned} M_h L_{(CA)} T K_{1,(E)} K_{2,(E)} K_{3,(E)} \\ = M_h L_{(CA)} T L_{(E)} \\ = M_h L_{(CA)} B \\ = Q. \end{aligned} \quad (15)$$

$$\begin{aligned} &\left(M_h L_{(CA)} T K_{1,(E)}\right)^{-1} \\ &= \left(M_h L_{(CA)} K_{3,(CA)}^{-1} K_{4,(CA)} K_{1,(CA)}^{-1} M_r K_{1,(E)}\right)^{-1} \\ &= \left(M_h M_r K_{1,(E)}\right)^{-1} \\ &= K_{1,(E)}^{-1} M_r^{-1} M_h^{-1}. \end{aligned} \quad (16)$$

### C. Security Evaluation and Discussions

The advantages of the proposed PQCMC are summarized and discussed as follows.

- The PQCMC implicit certificate includes solely a reconstruction value (akin to the public key of an end entity) and the end entity's information. No signature is necessitated within the implicit certificate.

- A random number $r$ is generated for each implicit certificate. Therefore, different implicit certificates can be attributed to the same end entity.

- The updated public key can be derived through the CA's public key, the reconstruction value in the end entity's implicit certificate, and the hash of both the CA's certificate and the end entity's certificate. This facilitates verification of the issuer of the implicit certificate without necessitating a signature.

- The proposed PQCMC supports the issuance of pseudonymous certificates. The updated public key is obtained by the reconstruction value $B$ within which a random invertible matrix $M_r$ is incorporated. As a result, the original public key of the end entity remains undisclosed.

### D. The Proposed Random Invertible Matrix

An efficient method is proposed for generating a random invertible matrix with a time complexity of $O(\zeta_1)$, as outlined in **Algorithm 1**. The inputs consist of a random number $r$ and the matrix size $\zeta_1$, which are used to generate an $\zeta_1 \times \zeta_1$ matrix $M_1$ and the inversed matrix $M_2$ of matrix $M_1$. The random number $r$ is adopted as the seed of a pseudo-random number generator, ensuring that the generated matrices exhibit no disparities when using the same random seed. The pseudo-random order can be generated and reflected in the matrix $I$. Subsequently, the $\zeta_1 \times \zeta_1$ matrices $M_1$ and $M_2$ can be derived based on the arrangement specified by the matrix $I$.

---

**Algorithm 1** Random invertible matrix generation method

**Input**: a random number $r$ and the matrix size $\zeta_1$
**Output**: a $\zeta_1 \times \zeta_1$ matrix $M_1$ and a $\zeta_1 \times \zeta_1$ matrix $M_2$
1: Set $r$ as the random seed
2: Create an 1 x $\zeta_1$ increment matrix $I$
3: Create an $\zeta_1 \times \zeta_1$ zero matrix $M_1$ and another $\zeta_1 \times \zeta_1$ zero matrix $M_2$
4: **for** $i = 0$ to $\zeta_1 - 1$ **do**
5:   Generate a pseudo-random integer number $j$ which is between 0 and $\zeta_1 - 1$
6:   Swap($I[i]$, $I[j]$)
7: **end for**
8: **for** $i = 0$ to $\zeta_1 - 1$ **do**
9:   Set $M_1[i][I[i]] = 1$
10:  Set $M_2[I[i]][i] = 1$
11: **end for**
12: **return** $M_1$ and $M_2$

## IV. EVALUATION AND DISCUSSIONS

The lengths and computation time are compared for evaluating the proposed PQCMC in the following subsections.

### A. The Comparison of Lengths

Table I offers a comparison of various McEliece methods [14],[18]-[22], focusing on the proposed PQCMC implicit certificate scheme. The objective is to illustrate the dimensions of the public key $L_{(E)}$ (with dimensions $\zeta_1 \times \zeta_2$), the reconstruction value $B$ (with dimensions $\zeta_2 \times \zeta_2$), and the signature $s$ (with dimensions $\zeta_2 \times 1$). Notably, the results indicate that the proposed PQCMC is capable of generating implicit certificates without requiring a signature, thereby obviating the need for the length of $s$. While the length of the reconstruction value in the proposed implicit certificate might exceed that of the public key in an explicit certificate, the proposed PQCMC enables the issuance of pseudonymous certificates to enhance privacy.

TABLE I. LENGTH COMPARISION

| McEliece Method | $(\zeta_1, \zeta_2)$ | The length of $L_{(E)}$ | The length of $B$ | The length of $s$ |
|---|---|---|---|---|
| [14] | (524, 1024) | 66 KB | 128 KB | 128 bytes |
| [19] | (1219, 1702) | 253 KB | 354 KB | 213 bytes |
| [20] | (1696, 2048) | 424 KB | 512 KB | 256 bytes |
| [21] | (1751, 2048) | 438 KB | 512 KB | 256 bytes |
| [22] | (2384, 3178) | 925 KB | 1233 KB | 397 bytes |
| [20] | (3604, 4096) | 1802 KB | 2048 KB | 512 bytes |
| [22] | (5208, 6944) | 4415 KB | 5886 KB | 868 bytes |

### B. The Comparison of Computation Time

The typical time complexity for generating a random invertible matrix is $O(n^3)$[23]. However, Kong et al. proposed an improved method based $g$ subgroups that reduces the time complexity to $o\left(\left(\frac{n}{g}\right)^3\right)$ [23], as shown in Table II. In this study, **Algorithm 1** achieves a time complexity of $O(n)$, enabling the efficient generation of a random invertible matrix.

TABLE II. THE TIME COMPLEXITY COMPARISION

| Random invertible matrix generation method | Time Complexity |
|---|---|
| [23] | $o\left(\left(\frac{n}{g}\right)^3\right)$ |
| The proposed method | $O(n)$ |

## V. CONCLUSIONS AND FUTURE WORK

The proposed PQCMC, serving as a post-quantum cryptography-based implicit certificate scheme, can issue pseudonymous certificates for enhancing privacy. However, a limitation of the proposed PQCMC lies in the length of the reconstruction value in V2X communication. To circumvent this limitation, a potential approach is to preload the implicit certificate onto devices and embed the certificate's digest into the secure protocol data unit, resulting in reduced lengths.